\documentclass[11pt]{article}
\usepackage{textcomp}
\usepackage{pdfsync}
\usepackage{fancyhdr}
\usepackage{amssymb}
\usepackage{amsmath}
\usepackage{hyperref}

\usepackage{graphicx}
\usepackage{latexsym}
\usepackage{appendix}
\usepackage{srcltx}
\def\centeron#1#2{{\setbox0=\hbox{#1}\setbox1=\hbox{#2}\ifdim
   \wd1>\wd0\kern.48\wd1\kern-.48\wd0\fi
   \copy0\kern-.48\wd0\kern-.48\wd1\copy1\ifdim\wd0>\wd1
   \kern.48\wd0\kern-.48\wd1\fi}}

\newcommand{\beq}{\begin{equation}}
\newcommand{\eeq}{\end{equation}}
\newcommand{\bea}{\begin{eqnarray}}
\newcommand{\eea}{\end{eqnarray}}
\newcommand{\ba}{\begin{array}}
\newcommand{\ea}{\end{array}}

\begin{document}

\hskip3cm

\vskip3cm

\begin{center}
 \LARGE \bf  Non-Fefferman-Graham asymptotics\\
 and holographic renormalization in New Massive Gravity
\end{center}

\vskip2cm

\centerline{\Large \Large Colin Cunliff$^{1}$}

\hskip2cm

\begin{quote}
Department of Physics, University of California-Davis, Davis, CA, US$^{1}$
\end{quote}

\hskip2cm

\vskip2cm

\centerline{\bf Abstract} The asymptotic behavior of new massive gravity (NMG) is analyzed for all values of the mass parameter satisfying the Breitenlohner-Freedman bound.  The traditional Fefferman-Graham expansion fails to capture the dynamics of NMG, and new terms in the asymptotic expansion are needed to include the massive graviton modes.  New boundary conditions are discovered for a range of values $-1<2m^2l^2<1$ at which non-Einstein modes decay more slowly than the Brown-Henneaux boundary conditions.  The holographically renormalized stress tensor is computed for these modes, and the relevant counterterms are identified up to unphysical ambiguities.\\
\underline{\hskip12cm}\\
${}^{1}$cunliff@physics.ucdavis.edu~~

\thispagestyle{empty}
\renewcommand{\thefootnote}{\arabic{footnote}}
\setcounter{footnote}{0}

\newpage

\section{Introduction}
Higher derivative extensions of general relativity have recently been the focus of much attention.  String theory and other quantum gravity models generically predict the existence of such terms, and they generally improve the renormalizability of the theory.  This increased focus is also partly motivated by recent models of low-energy modifications of general relativity that could provide an alternative to dark energy \cite{Hiterbichler:2011}.  However, analysis of the dynamics of such theories is in general a difficult task, complicated by the nonlinearity of the equations of motion.  The situation is improved in lower-dimensional models, where the reduction in degrees of freedom simplifies the dynamics while retaining many of the properties of higher-dimensional models.

New massive gravity (NMG) is a particular three-dimensional model with a specific combination of curvature-squared terms in the action \cite{Bergshoeff:2009hq,Bergshoeff:2009aq}.  Generically, theories with curvature-squared terms contain massive spin-2 and massless ghost-like scalar modes; in NMG, however, the coefficients in the action are chosen in such a way that the scalar modes are excised from the theory \cite{Deser:2009pr}.  Unlike its cousin topologically massive gravity (TMG) \cite{Deser:1982,Deser:1981wh}, the theory is parity preserving, and was originally investigated as the non-linear completion of Fierz-Pauli theory.  NMG shares some features with TMG.  In particular, it admits anti-de Sitter (AdS$_3$) spacetime as a vacuum solution and permits a larger class of asymptotically anti-de Sitter solutions than Einstein gravity alone \cite{AyonBeat:2009ba,Oliva:2009ip,Clement:2009bh,Ahmedov:2010tg}.  Thus NMG formulated around an AdS background has proven a fruitful toy model for exploring the AdS/CFT correspondence \cite{Liu:2009bk,Liu:2009kc,Giribet:2009qz,Ghodsi:2010,Alishahiha:2010bw,Grumiller:2009sn,Sinha:2010,Perez:2011gs,Hohm:2010jc,Giribet:2010ed,Kwon:2012hr}.

NMG also shares with TMG the undesirable feature that massive gravitons and BTZ black holes appear with opposite sign energy, reigniting discussions about the consistency of 3D (topologically or new) massive gravity about an AdS$_3$ vacuum.  Discussions in TMG centered on the consistency of strong boundary conditions that (it was hoped) could truncate massive graviton ghosts and render the background stable \cite{Strominger:2008cg,Maloney:2009ck}.  To be specific, TMG possesses a critical point in parameter space at which both Brown-Henneaux \cite{Brown:1986nw} and relaxed log boundary conditions \cite{Henneaux:2009pw,Grumiller:2008es} seem acceptable, in the sense that they yield finite charges at infinity and preserve the asymptotic symmetries.  The theory with log boundary conditions is conjectured to be dual to a logarithmic conformal field theory, which is known to be non-unitary, while the theory with Brown-Henneaux boundary conditions is dual to a chiral CFT and is conjectured to be stable (see \cite{Maloney:2009ck} for a review).  NMG similarly possesses critical points in the space of parameters at which multiple boundary conditions are possible.  This discussion of boundary conditions in NMG is thus essential to conclude the stability of the theory and to establish which geometries contribute to the partition function of the quantum theory.

Early approaches to the question of appropriate boundary conditions used the linearized theory to identify and determine the consistency of possible boundary conditions \cite{Liu:2009bk,Liu:2009kc}.  Attention has primarily focused on two critical points at which novel solutions appear.  NMG possesses a chiral point, analogous to that of TMG, at which log deformations from BTZ are allowed \cite{AyonBeat:2009ba,Clement:2009bh}.  At another critical point, new type black holes \cite{Oliva:2009ip}, characterized by a kind of gravitational `hair,' have been found that require a different relaxation of the Brown-Henneaux boundary conditions.  However, beyond these two critical points, only the Brown-Henneaux boundary conditions have been investigated and shown to be consistent at all points in parameter space \cite{Liu:2009kc}.  The possible relaxation from Brown-Henneaux at non-critical values has not been explored.

The Fefferman-Graham expansion \cite{Fefferman:1985ci} provides a natural tool for determining the asymptotic behavior of the metric and for addressing the question of appropriate boundary conditions for asymptotically (locally) anti-de Sitter spacetimes.  In Einstein gravity, it has already proven to be an important tool in holographic renormalization (see \cite{Skenderis:2002wp} for a review), and in computation of correlation functions in the boundary CFT.  Recently, it has been applied to NMG at a critical value of the coupling for the purpose of constructing the renormalized boundary stress tensor \cite{Kwon:2012hr}.  However, the expansion used in \cite{Kwon:2012hr} applies only at that particular point in parameter space, and the authors point out that the generic asymptotic expansion remains unknown.

This paper explores the asymptotic expansion of the metric at all values of the mass parameter in NMG.  After covering the basics of NMG, I review the definition of asymptotically anti-de Sitter spacetimes and the derivation of the Fefferman-Graham expansion, drawing particular attention to those steps that rely on the bulk equations of motion.  These are the steps at which the derivation of the asymptotic expansion in NMG diverges from that of Einstein gravity.  The traditional Fefferman-Graham expansion is then applied to NMG and shown to be insufficient for recovering the non-Einstein solutions, except at a few special points in the parameter space.  The next section introduces a modified asymptotic expansion that captures both Einstein and non-Einstein solutions.  All known exact asymptotically AdS solutions are shown to have asymptotic behavior given by this modified expansion.  This approach correctly identifies the weakened asymptotics at the critical points found in previous studies and also finds new regions of parameter space at which the massive non-Einstein modes obey weaker-than-Brown-Henneaux fall off.  The Brown-York boundary tensor is constructed for these modes in the parameter range $-1<2m^2<1$.  Several possible counterterms are considered, and the renormalized boundary tensor is obtained up to unphysical ambiguities.  The central charge of the dual CFT is determined by the trace of the renormalized stress tensor.  The final section summarizes the results, with some comments on implications and future steps.


\section{Setup and Equations of Motion}
The bulk action of new massive gravity (NMG) is given by \cite{Bergshoeff:2009hq,Bergshoeff:2009aq,Townsend:2009}
\begin{equation}\label{NMGaction}
S =\frac{\xi}{2\kappa^2}\int d^3x\sqrt{-g}\left[ \sigma R +
2\lambda + \frac{1}{m^2}K  \right]\,,
\end{equation}
where $2\kappa^2=16\pi G$, and the constants $\xi$ and $\sigma$ are introduced to control the overall sign of the action and the Einstein-Hilbert piece and take the values $\pm 1$.  In addition to the gravitational constant $G$ and the cosmological parameter $\lambda$, NMG contains the mass parameter $m^2$ of mass dimension two.

The new tensor $K$ is a specific combination of curvature squared terms defined by
\beq
 K = R_{\mu\nu}R^{\mu\nu} -\frac{3}{8}R^2\,.
\eeq
%
%
%
%
%
The equations of motion given by variation of the action with respect to the metric are
\begin{equation}\label{eom}
\xi \left[
 \sigma G_{\mu\nu} - \lambda g_{\mu\nu} + \frac{1}{2m^2}K_{\mu\nu}\right]=0\,,
\end{equation}
where
\begin{equation}
 K_{\mu\nu} = 2 \Box R_{\mu \nu} - \frac{1}{2}\nabla_\mu \nabla_\nu R - \frac{1}{2}\Box R g_{\mu \nu} +
    4R_{\mu \alpha \nu \beta}R^{\alpha \beta} - \frac{3}{2}RR_{\mu \nu} -
    R_{\alpha \beta}R^{\alpha \beta} g_{\mu \nu} + \frac{3}{8}R^2 g_{\mu \nu}.\label{Ktensor}
\end{equation}
I will allow both positive and negative values of $m^2$ and consider both signs of the Einstein-Hilbert action; however, $\xi$ will be set to unity from this point forward.

NMG admits an AdS$_3$ vacuum with effective cosmological constant $\Lambda$ related to the bare cosmological parameter $\lambda$ by
\beq \label{rel}
  \Lambda = 2m^2 \left[\sigma \pm \sqrt{1-\frac{\lambda}{m^2} }\right]\,.
\eeq
The AdS radius is given by $\Lambda = -\frac{1}{L^2}$.  It will turn out convenient to
use the effective cosmological constant and AdS radius through the remainder of the paper.

\section{Asymptotically AdS Spacetimes}
This section outlines the ingredients that go into the derivation of the Fefferman-Graham expansion and points out the steps that explicitly rely on the bulk equations of motion.  These are the steps where the derivation for the correct asymptotic expansion of new massive gravity diverges from that of Einstein gravity.  The third subsection explicitly demonstrates the failure of the Fefferman-Graham expansion to capture the full dynamics of the theory, except at a few critical values of the mass parameter.

\subsection{Conformally Compact Manifolds}
Consider a $(d+1)$-dimensional manifold with metric $(M,g)$, where $M$ is the interior of a manifold-with-boundary $\bar{M}$, and the bulk metric $g$ becomes singular on the boundary, denoted $\partial M$.  Suppose the existence of a smooth, non-negative defining function $z$ on $\bar{M}$ such that $z(\partial M)=0$, $dz(\partial M)>0$, and $z(M)>0$.  This can be used to define a non-degenerate metric on $\bar{M}$,
\beq
    \bar{g} = z^2 g.
\eeq
Then, in the language of Penrose \cite{Penrose:1986sa}, the pair $(M,g)$ is labeled conformally compact, and the choice of defining function determines a particular conformal compactification of $(M,g)$, with boundary located at $z=0$.

The metric $\bar{g}$ induces a metric $g_{(0)}$ on the boundary $\partial M$.  However, this metric is not unique, as a different defining function conformally rescales the boundary metric.  The bulk metric $(M,g)$ thus induces a conformal structure $(\partial M, [g_{(0)}])$ on the boundary, where $[g_{(0)}]$ denotes a conformal class of metrics.

The connection to anti-de Sitter space becomes apparent in the expansion of the curvature tensor (of the bulk metric) in powers of $z$, yielding
\beq
    R_{\mu \nu \rho \sigma} = -\bar{g}^{\alpha \beta} \nabla_\alpha z \nabla_\beta z \left(g_{\mu \rho}
        g_{\nu \sigma} - g_{\mu \sigma}g_{\nu \rho} \right) + \mathcal{O}\left(z^{-3}\right)
\eeq
Note that $\bar{g}^{-1}$ is of order $z^{-2}$, so the leading term is of order $z^{-4}$.  If, in addition, there exists a defining function that, to leading order, satisfies $\bar{g}^{\alpha \beta} \nabla_\alpha z \nabla_\beta z = \frac{1}{L^2}$, then the manifold $M$ is called asymptotically locally anti-de Sitter, in the sense that the curvature tensor approaches that of AdS with radius $L$ near the boundary.  Note that no restriction has been placed on the boundary topology.

\subsection{Fefferman-Graham Expansion}
On the manifold-with-boundary $\bar{M}$, using the defining function $z$ as one of the coordinates allows us to bring the metric $\bar{g}$ to Gaussian normal form near the boundary,
\beq
    \bar{g}_{\mu \nu}dx^\mu dx^\nu = d\bar{s}^2 = dz^2 + g_{ij}dx^i dx^j.
\eeq
where Greek indices run over $d+1$ dimensions, and Latin indices run over the $d$ non-radial coordinates.  Then the bulk (physical) metric becomes
\beq
  ds^2=z^{-2}\left(dz^2+g_{ij}dx^i dx^j \right). \label{poincare}
\eeq
The $d$- dimensional metric induced on a hypersurface of constant $z$ can be expanded in powers of $z$
\beq
  g_{ij}(z,x^k)=g^{(0)}_{ij}(x^k) + \cdots, \label{generalexpansion}
\eeq
where the subleading terms vanish on the boundary as $z\rightarrow 0$.

The specific form the expansion depends on the bulk theory.  Fefferman and Graham \cite{Fefferman:1985ci} first derived the asymptotic expansion for general relativity in $d+1$ dimensions
\beq
  g_{ij}(r,x^k) = g^{(0)}_{ij} + z^{2}g^{(2)}_{ij} + \cdots + z^{d}\left(g^{(d)}_{ij} + \ln z h^{(d)}_{ij}\right) + \cdots \label{GRexpansion}
\eeq
They found that coefficients of odd powers of $z$ vanish, and subleading terms $g^{(2k)}$, $2k<d$, are fixed by the boundary metric $g^{(0)}$.  However, only part of $g^{(d)}$ is determined by the boundary.  Specifically, the trace and covariant divergence of $g^{(d)}$ are solved in terms of the boundary metric, leaving other components free.  The ``log'' term $h^{(d)}$ is present only in even dimensions $d>2$ and is given by the metric variation of the conformal anomaly.  In three dimensions, the expansion truncates at fourth order, $g=g^{(0)}+z^2g^{(2)}+z^4g^{(4)}$, and the metric can be found exactly \cite{Skenderis:2000qe}.

However, these results are all consequences of the equations of motion and therefore are specific to Einstein gravity.  In general, different bulk theories lead to different asymptotic expansions.  Even with a gravitational action given by the Einstein-Hilbert action, gravity coupled to other fields can yield different expansions.  For example, three dimensional Einstein gravity coupled to a free massless scalar field is of the form (\ref{GRexpansion}) with a non-zero log term $h^{(2)}$ \cite{Solodukhin:2001}.  In this case, $h^{(2)}$ is determined by the boundary values of the fields and does not indicate a new degree of freedom of the metric.  Other potentials for a scalar field can yield odd powers of $z$ \cite{Berg:2002ae} and even log-squared terms \cite{Kanitscheider:2006ha} in the asymptotic expansion.

Higher derivative gravitational theories can have even more exotic behavior.  For example, topologically massive gravity (TMG) at the critical point $\mu \ell=1$ can also have the log term $h^{(2)}$ in the asymptotic expansion \cite{Henneaux:2009pw,Grumiller:2008es,Grumiller:2008ii,Henneaux:2010mo}.  However, several points distinguish this case from that of general relativity coupled to a free massless scalar.  In that example, the log term is fixed in terms of the boundary fields.  But in critical TMG, one component of the log term is unconstrained by the equations of motion.  Its inclusion the asymptotic expansion is allowed but not required, and this phenomena is related to the presence of new bulk degrees of freedom in the metric.  This apparent freedom in the asymptotic expansion initiated a vigorous discussion in the literature about appropriate boundary conditions, with some arguing that different boundary conditions yield different theories \cite{Maloney:2009ck}.  However, another approach motivated by AdS/CFT argued that the bulk theory determines the correct asymptotic expansion \cite{Skenderis:2009nt}.

Another unusual feature of critical TMG is that the log term in TMG is only available at the critical point $\mu \ell=1$.  At all other values of the mass parameter, the equations of motion force the log term to vanish. So, unlike the case of GR coupled to a free massless scalar, the asymptotic expansion of TMG depends not just on the form of the bulk action but also \emph{on the parameters}.  The asymptotic expansion of TMG at non-critical values of the mass parameter was examined in \cite{Cunliff:2011tm}, and the expansion was found to contain terms of the form $z^{n}$, where the exponent $n$ is a function of the mass parameter.  This is reminiscent of Einstein gravity coupled to a massive scalar with higher-than-quadratic polynomial potential \cite{Henneaux:2007ab}.  In that case, the asymptotic behavior of both the metric and scalar is dependent on the value of the mass, with the most relaxed asymptotics occurring when the scalar saturates the Breitenlohner-Freedman bound \cite{Breitenlohner:1982}.  So the appearance of mass-dependent asymptotic expansion in higher-derivative gravity, though surprising, is not unique.

New massive gravity exhibits some features similar to topologically massive gravity.  As in TMG, NMG possesses a critical point $2m^2=-\sigma$ at which the log term $h^{(2)}$ is allowed but not required, and either choice of boundary conditions seems acceptable.  Also, as in TMG, this term vanishes by the equations of motion at non-critical values.  NMG also possesses another critical point $2m^2=+\sigma$ at which interesting new solutions have been found.  In particular, new type black holes \cite{Bergshoeff:2009aq,Oliva:2009ip} have been discovered with relaxed asymptotics containing odd powers of the radius
\beq
  g_{ij} = g^{(0)}_{ij}+zg^{(1)}_{ij}+z^{2}g^{(2)}_{ij} + z^{3}g^{(3)}_{ij}+ \cdots
\eeq
Here, too, this expansion only works at this particular point in the space of parameters.  At generic values of the mass parameter, the coefficients of the odd-powered terms vanish.  Also, similar to the log term at $2m^2=-\sigma$, the subleading term $g^{(1)}$ is \emph{allowed} but not required, i.e. it is not determined by the boundary metric but results from the increased bulk degrees of freedom of the metric.  Additionally, the theory at $2m^2=+\sigma$ allows a new log term \cite{AyonBeat:2009ba,Grumiller:2010tj} $g=g^{(0)}+z \ln z h^{(1)} + z g^{(1)} + \cdots$.
%

To date, the asymptotic behavior of NMG at generic values of $m^2$ remains unknown.  That odd-powered and log asymptotics are allowed only at critical values of $m^2$ hints at the prospect that the correct asymptotic expansion, as in TMG, depends on the value of the mass parameter.

\subsection{Failure of the ``Traditional'' Fefferman-Graham Expansion}
Before investigating the possibility of a parameter-dependent expansion, I first demonstrate the need for a modified expansion by explicitly showing the failure of the ``traditional'' Fefferman-Graham expansion to capture the dynamics of NMG at all but a few critical values of the mass parameter.  As noted in the previous section, this failure is unsurprising, since the Fefferman-Graham expansion was originally derived for metrics satisfying the Einstein equations.

First note that the coordinate system employed in (\ref{poincare}) covers only part of the boundary, and it is more convenient to work in a global coordinate system.  The radial coordinate transformation $z=e^{-r/L}$ moves the boundary to infinity, and the expansion for three dimensional general relativity becomes
\begin{eqnarray}
    ds^2 &=& \frac{dr^2}{L^2} + \gamma_{ij} dx^i dx^j \nonumber \\
    \gamma_{ij} &=& e^{2(r/L)}\left(g^{(0)}_{ij} + e^{-2(r/L)}g^{(2)}_{ij}+e^{-4(r/L)}g^{(4)}_{ij} + \cdots\right). \label{GRFGexpansion}
\end{eqnarray}
For simplicity, the AdS radius is fixed at $L=1$, since it can always be reinstated later through dimensional analysis.

The technique now is to plug this expansion into the equations of motion and solve order by order.  For the moment, consider the expansion (\ref{GRFGexpansion}) with the log term $re^{-2r}h^{(2)}$ excluded.  This should not affect the results, as previous work has shown the log term to be consistent only at the chiral point $2m^2=+\sigma$, and the goal here is the solution at generic values of $m^2$.  In general, the boundary metric $g^{(0)}$ is a free field and is not fixed by the equations of motion.  However, to simplify the expansion, attention is restricted to solutions which asymptote to exact AdS$_3$ with light-cone coordinates on the boundary, i.e. the boundary metric is chosen so $g^{(0)}_{+-}=-1$ with diagonal components vanishing.\footnote{Note the distinction between ``asymptotically locally anti-de Sitter'' (AlAdS) spacetimes, in which the boundary metric is treated as a free field and the curvature tensor approaches that of anti-de Sitter space near the boundary, and ``asymptotically anti-de Sitter'' (AAdS) spacetimes, in which the metric asymptotes to the exact AdS metric at the boundary.  In a slight abuse of notation, I use AAdS to refer to both cases, though the context should make clear which notion is appropriate.}  Gauge-independent equations are given in the appendix.

In general relativity, the second order equations of motion fix the trace and divergence of $g^{(2)}$, and the other components are undetermined.  This shows up in the vanishing of the $\{ij\}$ equations of motion.  In new massive gravity, it is also true that the $\{ij\}$ equations vanish identically, and only the $\{rr\}$ and $\{ri\}$ equations restrict the metric.  In new massive gravity, these equations become
\bea
  \left(2\sigma - \frac{1}{m^2} \right) g_{+-}^{(2)}=0	\\			
  \left(\sigma + \frac{1}{2m^2}\right) \partial_- g_{++}^{(2)} = \left(\sigma - \frac{1}{2m^2}\right) \partial_+ g_{+-}^{(2)}\\
  \left(\sigma + \frac{1}{2m^2}\right) \partial_+ g_{--}^{(2)} = \left(\sigma - \frac{1}{2m^2}\right) \partial_- g_{+-}^{(2)}
\eea
%
These are the same restrictions on $g^{(2)}$ that appear in Einstein gravity, multiplied by a pre-factor dependent on $\sigma$ and $m^2$.  These equations exhibit the two critical points $2m^2=\pm \sigma$ that have previously been explored in the literature.  At the critical point $2m^2=+\sigma$, the constraint on $g^{(2)}_{+-}$ vanishes, while the off-diagonal components have the same constraints as in Einstein gravity.  In a gauge-independent language, the constraints on the divergence of $g^{(2)}$ are maintained, while the constraint on the trace disappears.  Conversely, at the chiral point $2m^2=-\sigma$, the constraint on the trace of $g^{(2)}$ holds but the constraints on the divergence of $g^{(2)}$ vanish.  These solutions will be explored in more detail in the next section; however, the solution at generic values of the mass parameter is just the Einstein solution
\beq
    g^{(2)}_{++} = L(x^+) \qquad \qquad g^{(2)}_{--} = \bar{L}(x^-).
\eeq
The program continues by plugging in the generic solution for $g^{(2)}$ and solving for $g^{(4)}$.  The fourth order equations generically constrain $g^{(4)}$ in terms of $g^{(2)}$ and $g^{(0)}$, but of course there are now three sets of equations: one each for the critical points $2m^2=\pm \sigma$, and the generic set using the second-order solution above.  The equations for generic $m^2$ are
\bea
    \{rr\}:&	\left(\sigma - \frac{1}{2m^2}\right)\left(4g_{+-}^{(4)} + g_{++}^{(2)} g_{--}^{(2)}\right)=0\\
    \{x^+x^+\}:&	\left(\sigma + \frac{17}{2m^2}\right) g_{++}^{(4)} = 0\\
    \{x^-x^-\}:&	\left(\sigma + \frac{17}{2m^2}\right) g_{--}^{(4)} = 0\\
    \{x^+x^-\}:&	\text{same as rr-eqn}\\
    \{rx^+\}:&	\left(\sigma + \frac{17}{2m^2}\right) \partial_- g_{++}^{(4)} = \left(\sigma - \frac{1}{2m^2}\right) \partial_+
        \left(g_{+-}^{(4)} + \frac{1}{4} g_{++}^{(2)} g_{--}^{(2)}\right)\\
    \{rx^-\}:&	\left(\sigma + \frac{17}{2m^2}\right) \partial_+ g_{--}^{(4)} = \left(\sigma - \frac{1}{2m^2}\right) \partial_-
        \left(g_{+-}^{(4)} + \frac{1}{4} g_{++}^{(2)} g_{--}^{(2)}\right)
\eea
Only the first three equations are necessary to solve for $g^{(4)}$, with the $\{x^+x^-\}$ duplicating the $\{rr\}$ equation, and the $\{rx^i\}$ equations being derivatives of combinations of the other equations.  This system has a new critical value $2m^2 = -17\sigma$ at which some of the constraints on $g^{(4)}$ vanish.  The vanishing of constraints is related to the additional degrees of freedom of the metric in NMG.  However, at generic values of the mass parameter, the only non-zero component is
\beq
    2m^2 \neq \pm \sigma, -17\sigma: \qquad g^{(4)}_{+-} = -\frac{1}{4}g^{(2)}_{++}g^{(2)}_{--} = -\frac{1}{4}L(x^+)\bar{L}(x^-)
\eeq
%
%
%
%

At sixth order, the same pattern emerges.  There are now four sets of equations: one each for the critical values $2m^2= \pm \sigma, -17\sigma$ and a generic set.  Again, the generic equations are just the Einstein equations multiplied by some pre-factor:
\bea
    \{rr\}:&	\left(\sigma - \frac{1}{2m^2}\right) g_{+-}^{(6)}=0\\
    \{x^+x^+\}:&	\left(\sigma + \frac{49}{2m^2}\right) g_{++}^{(6)}=0\\
    \{x^-x^-\}:&    \left(\sigma + \frac{49}{2m^2}\right) g_{--}^{(6)}=0 
\eea

As is the case for the fourth-order equations, the $\{x^+x^-\}$ equation duplicates the $\{rr\}$ equation, and the $\{rx^i\}$ equations are just derivatives of the other equations. These equations exhibit a new critical value $2m^2=-49\sigma$ at which some of the constraints on $g^{(6)}$ vanish.  However, at generic values of the mass parameter, all components of $g^{(6)}$ vanish.  It seems reasonable to assume that (at generic values of the mass parameter) all higher-order terms vanish as in Einstein gravity, and this has been confirmed to tenth order.  Then at generic values of the mass parameter, the traditional Fefferman-Graham expansion (\ref{GRFGexpansion}) truncates at fourth order and contains only the same solutions allowed in Einstein gravity.

Clearly, the expansion (\ref{GRFGexpansion}) fails to capture the dynamics of new massive gravity.  If (\ref{GRFGexpansion}) is the correct asymptotic expansion for NMG, these results suggest that the theory at generic values of $m^2$ contains only the ordinary Einstein solutions, but special values yield more degrees of freedom.  However, this conflicts with earlier work.  Perturbative methods have shown that new massive gravitons of NMG exist at a wide range of values of $m^2$.  Non-perturbative degree of freedom counting techniques find the same number of degrees of freedom at all values of the mass parameter \cite{Blagojevic:2010ir,Blagojevic:2011eg,Hohm:2012ot}.  Additional counter-examples exist in the literature.  For example, the AdS pp-waves of Ayon-Beato \cite{AyonBeat:2009ba}, and their generalization to the Type N solutions \cite{Ahmedov:2010tg,Ahmedov:2010em}, are exact solutions that exist at all values of the mass parameter.  For the range $2m^2\leq -\sigma$, the AdS waves are asymptotically anti-de Sitter and satisfy the strict Brown-Henneaux boundary conditions.  However, these solutions can not be put into the form (\ref{GRFGexpansion}), except at the critical values mentioned earlier.  These solutions will be discussed in greater depth in the next section.

Taken together, these results indicate that (\ref{GRFGexpansion}) is not the correct asymptotic expansion for new massive gravity at generic values of $m^2$.  Inclusion of the log term $h^{(2)}$in (\ref{GRFGexpansion}) will not help, since that term is only non-zero at $2m^2=-\sigma$.  Similarly, inclusion of odd terms $g^{(1)}, \ldots$ will not remedy this problem, as $g^{(1)}$ is allowed only at the critical point $2m^2=\sigma$.

Note again that this failure of the Fefferman-Graham expansion at generic values has been obscured in the literature, since previous studies exploring the asymptotic expansion of NMG have focused on the critical values $2m^2=\pm \sigma$ at which the FG expansion \emph{does} capture the new degrees of freedom of the theory.

\section{A Modified Asymptotic Expansion for NMG}
\subsection{Modified Asymptotics and Solution at Generic $m^2$}
The goal of this section is to remedy the failure of the Fefferman-Graham expansion noted in the previous section, which is accomplished by adding new terms to the asymptotic expansion.  Note that, regardless of the value of $m^2$, solutions of Einstein gravity are also solutions of new massive gravity.  Therefore, the generic expansion must still include the original terms of the FG expansion.  Another way of seeing this is that the constraints that fix the trace and divergences of $g^{(2)}$ in terms of the boundary metric hold at all values of $2m^2\neq \pm \sigma$, and so $g^{(2)}$ is required in the generic expansion.  Similarly, the equations of motion fix $g^{(4)}$ in terms of $g^{(2)}$ except at $2m^2\neq -17\sigma$, so the fourth-order term is also present in the generic expansion.

But the Einstein terms are insufficient to capture the dynamics at generic values of the mass parameter.  To that end, I propose a new term in the expansion
%
%
\beq
  \gamma_{ij}(r,x^i)=e^{2r}\left(g^{(0)}_{ij} + e^{-2r}g^{(2)}_{ij} + e^{-4r}g^{(4)}_{ij} + e^{-nr}g^{(n)}_{ij} + \cdots\right) \label{NMGFGexpansion}
\eeq
for some exponent $n$, to be determined later.  For the moment, the expansion will be carried through in a generic gauge, without any particular choice of coordinate system or boundary metric.  Next, the equations of motion are expanded to first order in $n$:
\bea
    \{rr\}:& \left(\sigma - \frac{1}{2m^2}\right)  \frac{-n}{2} \text{Tr} g^{(n)}=0 \label{Tracegn}\\
    \{ij\}:& -n\left(\frac{n}{2}-1\right)\left[\sigma + \frac{1}{m^2} \left(n^2-2n+\frac{1}{2}\right)\right] g_{ij}^{(n)}
        + n\left(\frac{n}{2}-1\right)\left[\sigma + \frac{n(n-2)}{2m^2}\right] g_{ij}^{(0)} \text{Tr} g^{(n)} = 0 \label{gnEOM}\\
    \{ri\}:& -\frac{n}{2}\left[\sigma + \frac{1}{m^2} \left(n^2-2n+\frac{1}{2}\right)\right]\nabla^k g^{(n)}_{ki} + \frac{n}{2}
        \left[\sigma + \frac{n(n-2)}{2m^2}\right]\partial_i \text{Tr} g^{(n)} =0
\eea
The derivation is given in the appendix.

Away from the critical point $2m^2=+\sigma$, the $\{rr\}$ equation imposes the vanishing of the trace of $g^{(n)}$, regardless of the value of $n$.  The $\{ij\}$ equations are more interesting.  The constraints on the non-trace part of $g^{(n)}$ vanish whenever the pre-factor, a quartic polynomial in $n$, vanishes.  This pre-factor has four roots:
\beq
    n= 0, 2, 1 \pm \sqrt{\frac{1}{2}-\sigma m^2} \label{branches}
\eeq
%
%
The four roots correspond to four branches of solutions, which should be expected since the equations of motion are fourth order in derivatives of the metric.  The first two roots $n=0, 2$ are also present in the Einstein limit $m^2 \rightarrow \infty$ and match up with the terms $g^{(0)}$ and $g^{(2)}$ which are already present in the ordinary Fefferman-Graham expansion.  The next two roots $n_\pm = 1 \pm \sqrt{\frac{1}{2} - \sigma m^2}$ correspond to the non-Einstein solutions.  These are the pieces of the expansion, new to NMG, which capture the dynamics of the theory.  Now the connection between the asymptotic expansion and the linearized approach becomes apparent.  The exponential behavior of the non-Einstein solutions can also be written as $n=1\pm m_{eff}$, where $m^2_{eff}=\frac{1}{2}-\sigma m^2$ is the effective mass of the massive graviton modes found in \cite{Liu:2009bk,Liu:2009pha}.  Additionally, the AdS pp-waves found in \cite{AyonBeat:2009ba} satisfy the Klein-Gordon equation with the same effective mass.  Note that imposing Brown-Henneaux boundary conditions $n\geq 2$ amounts to choosing only the positive branch and restricting to the parameter range $2m^2 \leq -\sigma$.

This explains the behavior observed in the previous section.  There it was found that the Fefferman-Graham expansion, containing only even powers of $e^r$ in the expansion, yields only Einstein solutions \emph{except} at the special values $2m^2 = -\sigma, -17\sigma, -49\sigma, \cdots$.  At these values, some of the equations vanish, and only at these critical points does the Fefferman-Graham expansion contain non-Einstein solutions.  From (\ref{branches}), it becomes clear that these are precisely the values of the mass parameter at which the non-Einstein branch of solutions $n_+$ overlaps with the Einstein branch, i.e., these are the values at which $n_+$ is a positive even integer.

Restricting attention only to those solutions satisfying Brown-Henneaux boundary conditions, we see that larger values of the mass parameter correspond to steeper asymptotics.  This phenomena has an obvious interpretation.  The coupling $m^{-2}$ gives the relative weight between the Einstein and NMG contributions to the action.  As $m^2$ increases, an NMG perturbation from an Einstein background decreases in strength, and the new degrees of freedom are more localized in the interior.  Conversely, when $2m^2 = \sigma$, the NMG and Einstein contributions in the action have the same relative weight, and NMG perturbations from an Einstein background show up at the same order in the asymptotic expansion as the Einstein degrees of freedom.

Additionally, the general theory possesses two points at which two branches of solutions degenerate.  When this occurs, the set of solutions labeled by (\ref{branches}) fails to span the space of linearly independent solutions, and new log solutions appear.  At the chiral point $2m^2=-\sigma$, the negative branch degenerates with the boundary at $n=0$, and the positive branch degenerates with the $n=2$ Einstein term.  Here the generic solution is
\beq
	\gamma_{ij} = e^{2r}\left(rh^{(0)}_{ij} + g^{(0)}_{ij} + re^{-2r}h^{(2)}_{ij} + e^{-2r}g^{(2)}_{ij} + \cdots\right) \label{chiralExp}
\eeq
This is similar to what happens in topologically massive gravity at the chiral point, at which the new log term $h^{(2)}$ is allowed.

NMG also possesses another critical point at $2m^2=+\sigma$, which has no analogue in topologically massive gravity.  Here, the two non-Einstein branches degenerate with each other at $n_+=n_-=1$, and the generic solution is given by
\beq
	\gamma_{ij} = e^{2r}\left(g^{(0)}_{ij} + re^{-r}h^{(1)}_{ij} + e^{-r}g^{(1)}_{ij} + \cdots \right). \label{criticalExp}
\eeq
The next section compares known exact solutions from the literature to the asymptotic behavior of the modified asymptotic expansion (\ref{NMGFGexpansion}).

\subsection{Known Solutions in Fefferman-Graham Coordinates}
Only the sub-leading behavior of non-Einstein perturbations has been established.  Finding exact solutions is more difficult.  For generic values of the exponent $n>2$, the next term in the expansion occurs at order $n+2$ and contains mixing between the Einstein background and the non-Einstein perturbation.  However, self-interactions quadratic in $g^{(n)}$ occur at order $2n$, and a non-zero $g^{(n)}$ turns on an infinite series with exponential behavior $ln+2k$, where $k$ and $l$ are positive integers.  Thus a generic perturbation from an Einstein background seems to generate an expansion that continues indefinitely, making this a poor tool for finding exact solutions.

Still, exact asymptotically AdS solutions are known, and a test of the modified expansion is whether it can accommodate all known solutions.  BTZ black holes \cite{btz} exist at all values of the mass parameter, and this matches what was stated in the previous section.  At all values, Einstein solutions are also solutions of NMG, and the second- and fourth- order terms in the expansion are required at all values of the mass parameter.  These solutions results from turning the non-Einstein perturbations $g^{(n_\pm)}$ off.

Most new non-Einstein exact solutions exist at one of the critical points $2m^2 = \pm \sigma$, and these are addressed next.  Early on, log deformations with terminating FG expansion were found at the chiral point $2m^2=-\sigma$.  These require a non-zero $h_{ij}^{(2)}$ in the general expansion (\ref{chiralExp}).  Additionally, new hairy black holes where found at the critical point $2m^2=+\sigma$ which also require relaxed asymptotics, and result from turning on the $g_{ij}^{(1)}$ piece.  The static hairy black holes can be written in Fefferman-Graham coordinates as
\beq
    ds^2 = dr^2 - a\sinh^2 r dt^2 + \left(a \cosh r + c\right)^2 d\phi^2
\eeq
This solution has finite Fefferman-Graham expansion that terminates at $g^{(4)}$ and is thus contained in the general expansion (\ref{criticalExp}).  This solution can also be boosted to a rotating black hole that, however, does not spoil the asymptotics.  For more details on the hairy black holes, see \cite{Bergshoeff:2009aq,Oliva:2009ip,Kwon:2012hr,Kwon:2011ey,Giribet:2009qz,Giribet:2010ed}.

These solutions are not good tests of the generic solution because they occur only at critical points at which the Fefferman-Graham expansion works or requires only minimal modification such as with the inclusion of log terms or odd-exponent terms.  A better test is to compare the expansion at generic $m^2$ (\ref{NMGFGexpansion}) with non-Einstein AAdS solutions at non-critical points.  The only known solutions that fit the bill are the AdS waves \cite{AyonBeat:2009ba} and Type N solutions \cite{Ahmedov:2010tg,Ahmedov:2010em}, both of which exist at all values of the mass parameter.  AdS waves are a kind of exact gravitational wave, conformally related to pp-waves, propagating on an AdS background with a null Killing vector.  In Fefferman-Graham coordinates, the AdS wave solutions can be written in light cone gauge as
\beq
    ds^2 = dr^2 - 2e^{2 r} dx^+dx^- + F_\pm(x^+) e^{(2-n_\pm)r} dx^+dx^+ \label{AdSwaves}
\eeq
where $F$ is unconstrained by the equations of motion.  The exponent $n$ is exactly the function of $m^2$ found for the generic perturbation, $n_\pm=1 \pm \sqrt{1/2-\sigma m^2}$.

Surprisingly, the expansion (\ref{NMGFGexpansion}) also accommodates exact solutions that are not asymptotically anti-de Sitter.  The Type N solutions of Ahmedov and Aliev \cite{Ahmedov:2010em,Ahmedov:2010tg} are exact solutions with constant scalar curvature that extend ``beyond the boundary.''  These are generalizations of the AdS waves (\ref{AdSwaves}) in which the traceless part of the Ricci tensor can be written as the exterior product of a null vector with itself.  The metric for these solutions is given by
\beq
    ds^2 = dr^2 + 2\cosh^2 r dx^+ dx^- + \left[Z(x^+,r)- (x^-)^2 \cosh^2 r \right]dx^+ dx^+ \label{TypeN}
\eeq
where
\beq
    Z(x^+,r)=F_1(x^+)\left(e^{(2-n_-)r} + e^{-n_-r}\right) + F_2(x^+)\left(e^{(2-n_+)r} + e^{-n_+r}\right) + F_3(x^+)\left(e^{2r}-e^{-2r}\right)
\eeq
and the exponent is again given by $n_\pm = 1\pm \sqrt{1/2 - \sigma m^2}$.  These solutions were shown to reduce to the AdS waves in an appropriate limit.  While they have a more complicated expansion than the AdS waves, the first sub-leading terms are still of order $e^{(2-n_\pm)r}$.

Note that neither the AdS waves (\ref{AdSwaves}) nor the Type N metric (\ref{TypeN}) can be accommodated by the original Einstein Fefferman-Graham expansion, and these solutions can be taken as early indication that the full asymptotic expansion requires modification from that found for Einstein gravity.  Instead, the generic expansion (\ref{NMGFGexpansion}) with non-Einstein asymptotics determined by (\ref{branches}) gives the correct asymptotic behavior of all known exact AAdS solutions.

\subsection{Boundary Conditions}
The near-boundary asymptotic expansion employed in the previous section provides a natural setting to address the question of appropriate boundary conditions.  Early on, the Brown-Henneaux boundary conditions were shown to be consistent at all values of the mass parameter .  The log solutions (at the chiral point $2m^2=-\sigma$) and the new type black holes (at the critical point $2m^2=+\sigma$) spurred investigations into these relaxed boundary conditions, and they were also found to be consistent at these points in parameter space.  Consistency, in this case, means that asymptotic symmetries of AdS preserve the boundary conditions, and conserved charges, expressed as surface integrals at infinity, remain finite.

However, the possibility of relaxing the Brown-Henneaux boundary conditions has been addressed only at the critical points $2m^2=\pm \sigma$.  The solutions associated with the non-Einstein branches indicate several other possibilities for relaxing the boundary conditions at generic couplings.  The asymptotic behavior of these solutions can be divided into four categories,\footnote{For the moment, I restrict discussion to the case $\sigma=+1$.  The negative case can be easily identified from (\ref{branches}).} to be addressed separately.
\begin{enumerate}
	\item	The negative branch in the parameter range $2m^2<-1$, corresponding to exponent $n_-<0$ and the new term $e^{-n_-r}g^{(n_-)}_{ij}$ in the asymptotic expansion
	\item	The negative branch in the range $-1<2m^2<1$, corresponding to exponent $0<n_-<1$
	\item	The positive branch in the range $-1<2m^2<1$, with exponent $1<n_+<2$
	\item	The positive branch in the range $2m^2<-1$, with expansion exponent $n_+ > 2$
\end{enumerate}

The first category consists of solutions that extend ``beyond the boundary'' and violate AdS asymptotics.  While some known exact solutions make use of this branch, these solutions cannot properly be called ``asymptotically AdS''.  The second category consists of solutions which asymptote to the AdS metric at the boundary but have slower-than-Brown-Henneaux fall-off.  Note that these solutions also decay more slowly than the fall-off for the new type black holes examined in \cite{Kwon:2012hr}.

The positive branch of solutions in the parameter range $-1<2m^2<1$ also asymptote to the AdS metric at the boundary but break the Brown-Henneaux boundary conditions.  In this range, the metric has asymptotic expansion
\begin{equation}
    \gamma_{ij} = e^{2r}\left(g^{(0)}_{ij} + e^{-n_+r}g^{(n_+)}_{ij} + e^{-2r}g^{(2)}_{ij} + \cdots \right) \label{positiveExpansion}
\end{equation}
with exponent $1<n_+<2$.  Though weaker than Brown-Henneaux, these solutions decay faster than the $n=1$ boundary conditions at the critical point $2m^2=+1$, which have already been shown to be consistent.  This behavior offers the intriguing possibility of new boundary conditions in a previously unexplored region of parameter space.  A next step to determine the consistency of these boundary conditions is to tackle the subject of holographic renormalization in this parameter range, which is addressed in the next section.

The fourth category contains solutions which decay faster than Brown-Henneaux.  Because they appear deeper in the interior, they are unlikely to affect either the counter-terms necessary for holographic renormalization or the conserved charges.

\section{Holographic Renormalization with Relaxed Boundary Conditions}
Previously, investigation of appropriate boundary conditions has been limited to i) Brown-Henneaux boundary conditions at all values of the mass parameter \cite{Liu:2009kc}, and ii) relaxations of Brown-Henneaux at the critical points $2m^2=\pm \sigma$, and holographic renormalization of the boundary stress tensor has only been achieved under these conditions \cite{Hohm:2010jc,Giribet:2010ed,Alishahiha:2010bw,Kwon:2012hr}.  However, results from the previous section indicate other possibilities for relaxing the boundary conditions over a range of parameters, and I explore one of these possibilities here.  In particular, the $n_+$ branch of solutions in the range $-1<2m^2<1$ falls off slower than Brown-Henneaux but faster than the $e^{-r}$ fall-off that was previously found to be consistent \cite{Giribet:2010ed,Kwon:2012hr}.  In this section I obtain the Brown-York stress tensor \cite{Brown:1992br} with these relaxed asymptotics and determine the appropriate counterterms necessary for renormalization.

Note that there are two further possibilities that will not be explored.  The $n_-$ branch in the range $-1<2m^2<1$ asymptotes to the AdS metric but obeys weaker-than Brown-Henneaux boundary conditions.  However, the exponent falls in the range $0<n_-<1$, and the Brown-York stress tensor must be expanded at least to second order in $g^{(n_-)}$ to include all divergent terms,
\begin{equation}
    T^{BY}_{ij} = e^{2r} T^{(0)}_{ij} + e^{(2-n_-)r}T^{(n_-)}_{ij} + e^{(2-2n_-)r}T^{(2n_-)}_{ij} + \cdots + e^0 T^{(2)}_{ij} + \cdots
\end{equation}
This higher-order expansion is required for the negative branch because terms quadratic in $g^{(n_-)}$ are found at order $e^{(2-2n_-)r}$, which is also divergent.  This more difficult problem is postponed for future research.  Also, in the parameter range $2m^2<-1$, the $n_-$ branch extends ``beyond the boundary'' and breaks the asymptotic symmetries of anti-de Sitter space.  While a renormalized stress tensor can be obtained for some non-AAdS spacetimes (see \cite{Hohm:2010jc}), holographic renormalization for these cases will not be explored here.



\subsection{Brown-York Stress Tensor with Relaxed Asymptotics}
In the AdS/CFT dictionary, the expectation value of the stress-energy tensor of the dual CFT is given by the renormalized Brown-York stress energy tensor evaluated at the boundary.  This is just
\begin{equation}
  <T^{ij}_{\text{CFT}}> = T^{ij}_{\text{ren}} = \frac{2}{\sqrt{-\gamma}}\frac{\delta}{\delta \gamma_{ij}}
    \left(S_{\text{bulk}} + S_{GH} + S_{c.t.}\right)
\end{equation}
where $S_{GH}$ is the generalized Gibbons-Hawking term \cite{Gibbons:1976ue} necessary for a well-defined variational principle, and $S_{c.t.}$ is the counter-term required to cancel divergences.  Together, the first two terms constitute the Brown-York stress energy tensor.

I begin by reviewing the auxiliary tensor formulation used by Hohm and Tonni \cite{Hohm:2010jc}.  In this approach, the NMG action (\ref{NMGaction}) is written in terms of an auxiliary field $f_{\mu \nu}$:
\begin{equation}
  S=\frac{1}{2\kappa^2}\int d^3 x\sqrt{-g}\left[\sigma R + 2\lambda + f^{\mu \nu}G_{\mu \nu} - \frac{m^2}{4}
    \left(f^{\mu \nu}f_{\mu \nu} - f^2\right)\right]. \label{NMGauxaction}
\end{equation}
where $G_{\mu \nu} = R_{\mu \nu} - \frac{1}{2}Rg_{\mu \nu}$ is the Einstein tensor.  On-shell, the auxiliary tensor is proportional to the Schouten tensor
\begin{equation}
  f_{\mu \nu} = \frac{2}{m^2}\left(R_{\mu \nu} - \frac{1}{4}Rg_{\mu \nu}\right) \label{auxiliary}
\end{equation}
and the equation of motion is
\begin{eqnarray}
  \sigma G_{\mu \nu} + \lambda g_{\mu \nu} - \frac{m^2}{2}\left[f_\mu^\alpha f_{\nu \alpha} - f f_{\mu \nu} -
    \frac{1}{4}g_{\mu \nu}\left(f^{\alpha \beta}f_{\alpha \beta} - f^2\right)\right] & \nonumber \\
  + 2f_{\alpha(\mu}G^{\alpha}_{\nu)} + \frac{1}{2}Rf_{\mu \nu} - \frac{1}{2}fR_{\mu \nu} -
    \frac{1}{2}g_{\mu \nu}f^{\alpha \beta}G_{\alpha \beta} & \nonumber \\
  + \frac{1}{2}\left[D^2 f_{\mu \nu} - 2D^\alpha D_{(\mu}f_{\nu)\alpha} + D_{\mu} D_{\nu} f +
    \left(D^\alpha D^\beta f_{\alpha \beta} - D^2 f\right)g_{\mu \nu}\right] & =0
\end{eqnarray}

The first step in constructing the Brown-York stress energy tensor is finding the appropriate generalization of the Gibbons-Hawking term.  This is a boundary action added to (\ref{NMGauxaction}) in order to have a well-defined variational principle, such that variations of the metric vanish at the boundary but their normal derivatives do not.  For Einstein-Hilbert gravity, the Gibbons-Hawking term is just the trace of the extrinsic curvature, which in Gaussian normal coordinates\footnote{See \cite{Hohm:2010jc} for gauge-independent definitions.} is
\begin{equation}
    K_{ij} = -\frac{1}{2}\partial_r \gamma_{ij} \qquad \text{and} \qquad K = \gamma^{ij}K_{ij}.
\end{equation}
It is also useful to decompose the auxiliary tensor in radial and non-radial components:
\begin{equation}
    f^{\mu \nu} = \left(
                    \begin{array}{cc}
                      s & h^i \\
                      h^i & f^{ij} \\
                    \end{array}
                  \right)
\end{equation}
Then the generalized Gibbons-Hawking term for new massive gravity is just
\begin{equation}
    S_{GH} = \frac{1}{2\kappa^2}\int_{\partial M_3} \left(-2\sigma K - \hat{f}^{ij}K_{ij} + \hat{f}K\right).
\end{equation}
With this action, the Brown-York stress energy tensor for NMG was obtained in \cite{Hohm:2010jc,Giribet:2010ed},
\begin{eqnarray}
    8\pi GT_{BY}^{ij} &=& \sigma\left(K^{ij} - K\gamma^{ij}\right) + \frac{1}{2}(\hat{s}-\hat{f})(K^{ij}-K\gamma^{ij})
        -\nabla^{(i}\hat{h}^{j)} + \frac{1}{2}\mathcal{D}_r \hat{f}^{ij} \nonumber \\
    & & + K_k^{(i}\hat{f}^{j)k} + \gamma^{ij}\left(\nabla_k \hat{h}^k - \frac{1}{2}\mathcal{D}_r \hat{f}\right),
\end{eqnarray}
where the first term is just the Brown-York tensor for Einstein gravity.  Expressions for the ``covariant r-derivative'' $\mathcal{D}_r$ and hatted quantities are given in \cite{Hohm:2010jc}.  Note that in Gaussian normal coordinates, $\mathcal{D}_r$ becomes the ordinary $r-$ derivative $\partial_r$, and $\hat{\phantom{m}}$ has no effect: $\hat{f}=\gamma_{ij}f^{ij}$, $\hat{h}^i=h^i$, $\hat{s}=s$.

We are now in a position to expand the Brown-York stress tensor using the modified asymptotic expansion given in (\ref{positiveExpansion}).  To simplify notation, I am dropping the subscript from $n_+$.  Then
\begin{eqnarray}
    8\pi G T_{BY}^{ij} &=& e^{-2r}\left(\sigma + \frac{1}{2m^2}\right)g_{(0)}^{ij} \nonumber \\
    & & + e^{-(n+2)r}\left[
        \left(\frac{n-2}{2}\sigma + \frac{2n^3 - 4n^2 + n - 2}{4m^2}\right)g_{(n)}^{ij} +
        \left(\frac{-n}{2}\sigma + \frac{-n^3 + 2n^2 - 2n}{4m^2}\right)g_{(0)}^{ij}\text{Tr}g^{(n)}\right] \nonumber \\
    & & + e^{-4r}\left[\frac{-1}{4m^2}R^{(0)}g_{(0)}^{ij} - \left(\sigma + \frac{1}{m^2}\right)g_{(0)}^{ij} \text{Tr}g^{(2)}
        \right] + \mathcal{O}(e^{-(n+4)r}) \label{BYFGexpanded}
\end{eqnarray}
There are now two divergences coming from the first two terms in the stress-energy tensor.

\subsection{Counter-terms and Renormalized Stress Tensor}
In this section I construct the relevant counter-terms and obtain the renormalized stress tensor.  This stress tensor gives the correct central charge of the dual CFT, as well as the mass and angular momentum of BTZ black holes, and is consistent with previous results obtained by other methods.  Previous work has explored holographic renormalization at the chiral point $2m^2=-\sigma$ \cite{Hohm:2010jc,Alishahiha:2010bw} and the critical point $2m^2=+\sigma$ \cite{Giribet:2010ed,Kwon:2012hr}, and their results are briefly reviewed next.

Holographic renormalization at the chiral point $2m^2=-\sigma$ was first explored in \cite{Hohm:2010jc,Alishahiha:2010bw}.  At this point, the correct expansion (\ref{chiralExp}) corresponds to turning on the log branch $h^{(2)}$.  Then log divergences appear in the Brown-York stress tensor (\ref{BYFGexpanded}).  The counter-term required to remove both the leading order and sub-leading log divergences is just proportional to the boundary cosmological constant, as in Einstein gravity:
\begin{equation}
    S_{c.t.} = -\left(\sigma + \frac{1}{2m^2}\right)\frac{1}{8\pi G}\int d^2x \sqrt{-\gamma} \label{BHcounterterm}
\end{equation}
This counter-term is sufficient for solutions obeying Brown-Henneaux boundary conditions at all values of the mass parameter.  Interestingly, it is also sufficient to cancel the log divergence at the chiral point---no new counter-terms are needed.  This is reminiscent of what happens in topologically massive gravity at the chiral point.  The authors of \cite{Hohm:2010jc} explored the asymptotic symmetry algebra of the renormalized stress tensor and confirmed that it reproduces the correct central charge of the dual CFT.

At the critical point $2m^2=+\sigma$, the counter-term (\ref{BHcounterterm}) is insufficient to cancel divergences in the boundary stress tensor.  The full asymptotic expansion is given by (\ref{criticalExp}); however, previous work studying holographic renormalization at the critical point has considered only the solutions with the $g^{(1)}$ branch of solutions turned on \cite{Giribet:2010ed,Kwon:2012hr}.  The log branch associated with non-zero $h^{(1)}$ has not been considered in this context and remains an open question.\footnote{The theory with non-zero $h^{(1)}$ has been referred to as ``partially massless NMG'', and its properties have been explored from the perspective of the dual CFT in \cite{Grumiller:2010tj}.}  With a non-zero $g^{(1)}$, the Brown-York tensor (\ref{BYFGexpanded}) has a sub-leading divergence of order $e^r$.  The counter-term necessary for removing both leading and sub-leading divergences is
\begin{equation}
    S_{c.t.} = \frac{m^2}{2}\left(\sigma+\frac{1}{2m^2}\right){8\pi G} \int d^2 x \sqrt{-\gamma}\hat{f} \label{CriticalCounterTerms}.
\end{equation}

Note that NMG brings with it an expanded set of possible counter-terms, with the only criterion being that they be constructed from purely local objects that are invariant under boundary-preserving diffeomorphisms.  Here I consider the expanded set of counter-terms given in \cite{Kwon:2012hr}:
\begin{equation}
    S_{c.t.} = \frac{1}{8\pi G}\int d^2 x \sqrt{-\gamma}\left(A + B \hat{f} + C\hat{f}^2 + Df_{kl}f^{kl}\right) \label{GenCT},
\end{equation}
with coefficients fixed by the requirement that the renormalized stress tensor remain finite.  The expansion of the counter-terms is also given in the appendix.  The leading-order divergence is removed when
\begin{equation}
    \left(\sigma + \frac{1}{2m^2}\right) + A - \frac{2}{m^2}B + \frac{4}{m^4}C + \frac{2}{m^4}D = 0 \label{leadingDiv}
\end{equation}
At sub-leading order, there appear to be two divergences: one proportional to $g^{(n)}$ and the other proportional to $\text{Tr}g^{(n)}$, which can be removed by the counter-terms when:
\begin{eqnarray}
  \frac{n+2}{2}\sigma + \frac{2n^3-4n^2+n+2}{4m^2} + A - \frac{2}{m^2}B + \frac{4}{m^4}C + \frac{2}{m^4}D &=& 0 \label{subDiv}\\
  -\frac{n}{2}\sigma + \frac{-n^3+2n^2-2n}{4m^2} + \frac{n}{m^2}B - \frac{4n}{m^4}C - \frac{2n}{m^4}D &=& 0 \label{TrDiv}
\end{eqnarray}
However, recall that the on-shell equations of motion (\ref{Tracegn}) fix $\text{Tr}g^{(n)}=0$, and so the constraint coming from (\ref{TrDiv}) is not needed.  Also, recall that the exponent $n$ is not an independent parameter.  The term $g^{(n)}$ is only non-zero when $n=1+\sqrt{\frac{1}{2}-\sigma m^2}$.  With this value of $n$, the equation (\ref{subDiv}) reduces to (\ref{leadingDiv}), and removal of the $g^{(n)}$ divergence places no new restrictions on the counter-terms.  Oddly, \emph{either} counter-term (\ref{BHcounterterm}) (for Brown-Henneaux boundary conditions, with to $B=C=D=0$) or the critical point counterterm (\ref{CriticalCounterTerms}) (with $A=C=D=0$) is sufficient for removing divergences from the stress energy tensor.  Thus the requirement that counterterms remove divergences in the boundary stress tensor is not enough to fix the counterterm.  However, these ambiguities are unphysical, and any choice of $A, B, C, D$ satisfying (\ref{leadingDiv}) leads to the same renormalized stress tensor.

With divergences removed, the renormalized stress tensor is
\begin{eqnarray}
  8\pi G T^{ren}_{ij} & = & \left(\sigma + \frac{1}{2m^2}\right)g^{(2)}_{ij} + g^{(0)}_{ij}\text{Tr}g^{(2)}
    \left(-\sigma - \frac{1}{m^2} + \frac{2}{m^2}B - \frac{8}{m^2}C - \frac{4}{m^2}D\right) \nonumber \\
  & & + g^{(0)}_{ij}R^{(0)}\left(-\frac{1}{4m^2} + \frac{1}{m^2}B - \frac{4}{m^2}C - \frac{2}{m^2}D \right)
\end{eqnarray}
Recall the on-shell $\{rr\}$ equation of motion fixes
\begin{equation}
  R^{(0)}=-2\text{Tr}g^{(2)},
\end{equation}
and so the on-shell renormalized stress tensor becomes
\begin{equation}
  8\pi G T^{ren}_{ij} = \left(\sigma + \frac{1}{2m^2}\right)\left(g^{(2)}_{ij} - g^{(0)}_{ij} \text{Tr}g^{(2)}\right). \label{renormalizedStress}
\end{equation}
The trace anomaly can be written in terms of the boundary Ricci scalar
\begin{equation}
  8\pi G T = \frac{1}{2}\left(\sigma + \frac{1}{2m^2}\right)R^{(0)}
\end{equation}
which is consistent with the central charge
\begin{equation}
  c = \left(\sigma + \frac{1}{2m^2}\right) \frac{3}{2G}
\end{equation}
and reproduces results from \cite{Liu:2009bk,Hohm:2010jc}.

Several features differentiate these results from the case $2m^2=+\sigma$, and the results above do not hold smoothly in the limit $n\rightarrow 1$.  The divergent terms in the Brown-York tensor (\ref{BYFGexpanded}) give the correct expressions in this limit \cite{Kwon:2012hr}; however, the equation of motion forcing the trace of $g^{(n)}$ to vanish (\ref{Tracegn}) disappears in the limit $n\rightarrow 1$, and thus the Brown-York stress tensor has two independent constraints on the counter-terms needed to renormalize the stress tensor.  This explains why the counter-term found at $2m^2=+\sigma$ (\ref{CriticalCounterTerms}) differs from the counter-term found at the chiral point $2m^2=-\sigma$ (\ref{BHcounterterm}).  Second, the $g^{(1)}$ terms must be expanded to second-order and mix with the $g^{(2)}$ terms.  Thus, they modify expressions for the conserved charges.  However, the expression for the renormalized stress tensor (\ref{renormalizedStress}) makes it clear that, for generic $1<n<2$, the non-Einstein mode $g^{(n)}$ does not contribute to conserved charges.

This result is remarkably uninteresting.  It is surprising that the renormalized stress tensor, even with these relaxed asymptotics, is just proportional to the stress tensor for AdS$_3$ \cite{Balasubramanian:1999re}, and that the conserved charges of the non-Einstein modes vanish in this parameter range.  The results from this paper and other on holographic renormalization indicate that non-Einstein modes \emph{never} contribute to conserved charges \emph{except} at the critical points $2m^2 = \pm \sigma$.  At present, I do not have an explanation for why this should be the case.

\section{Conclusion}
In this paper, I have adopted the Fefferman-Graham approach to examine the asymptotic behavior of New Massive Gravity at generic couplings.  At generic values of the mass parameter, the traditional Fefferman-Graham expansion fails to capture the dynamics of the theory, demonstrating the need for a more general expansion.  The expansion at all values of the mass parameter is derived and used to find the asymptotic behavior of non-Einstein solutions to first order.  At the critical points $2m^2=\pm \sigma$, some of the branches of solutions degenerate, and new logarithmic solutions becomes possible at these values.  The validity of the general asymptotic expansion is confirmed by comparing to known exact solutions in the literature, and all known asymptotically AdS solutions are shown to match the asymptotic behavior found in this paper.

%
%
The solutions indicate a range of the mass parameter $-1<2m^2<1$ in which the non-Einstein solutions asymptote to the AdS metric with slower fall-off than Brown-Henneaux.  In particular, the positive branch solutions, though weaker than Brown-Henneaux, decay faster than the relaxed asymptotics at the critical point $2m^2=+\sigma$, which have previously been shown to be consistent.  Using the auxiliary tensor formulation of Hohm and Tonni, I have computed the Brown-York stress energy tensor with these asymptotics, and the appropriate counterterms required to remove divergences are determined up to unphysical ambiguities.  The holographically renormalized stress tensor is found and gives the correct central charge of the dual CFT.

However, the results are intriguingly vague.  First, the new divergences in the unrenormalized stress tensor stemming from the relaxed asymptotics cancel on-shell, with no new constraints on appropriate counterterms beyond those already needed for Brown-Henneaux boundary conditions.  Second, the renormalized stress tensor is exactly that of the theory with Brown-Henneaux boundary conditions, i.e. without the non-Einstein modes.  In other words, the non-Einstein solutions, despite the weaker fall-off, do not contribute to the boundary stress tensor or to the conserved charges.  More work is needed to interpret these results, especially in the context of the boundary CFT.

Another intriguing question is the possibility of holographic renormalization of non-AdS asymptotics.  Hohm and Tonni's original approach was applied successfully to asymptotically Lifshitz solutions, despite the fact that such solutions break the asymptotic symmetries of AdS spacetime.  The modified Fefferman-Graham expansion found here (\ref{NMGFGexpansion}) contains one branch of solutions -- the negative branch associated with $n_-$ in the parameter range $2m^2<-1$ -- that extend ``beyond the boundary'' and also break the asymptotic symmetries, and it would be interesting to see if the approach of \cite{Hohm:2010jc} can be used to define a finite stress tensor for these asymptotics.  The existence of exact solutions with this behavior, namely a subset of the AdS pp-waves and the Type N solutions, makes this question a potentially interesting one.

These results may also have some implications for recent investigations into critical gravity in higher dimensions.  Critical gravity \cite{Deser:2011} is the extension of the action (\ref{NMGaction}) to higher dimensions, formulated around the critical point $2L^2 m^2=\sigma (D-2)$, where $D$ is the (bulk) spacetime dimension and $L$ the AdS radius.  It was observed that ghost-like massive modes have asymptotic fall-off slower than massless modes, and the authors of \cite{Lu:2011cg} speculated that if these massive modes could be truncated by appropriate boundary conditions, the theory would be classically equivalent to Einstein gravity. This work was more recently extended to non-critical gravity \cite{Lu:2011,Hyn:2012ne}, where the slower fall-off of massive modes was observed for a range of couplings
\beq
    \frac{D^2-6D+7}{4(D-2)L^2}<m^2<\frac{D-2}{2L^2} \label{noncriticalmass}.
\eeq
Most of these studies are based on the linearized equations of motion; however, the modified Fefferman-Graham approach employed here offers a natural way to formulate questions concerning boundary conditions and could complement the linearization studies.  For $D=3$, (\ref{noncriticalmass}) is precisely the range at which both the positive and negative branches $n_\pm$ of non-Einstein solutions i) asymptote to the AdS metric at infinity, and ii) have weaker than Brown-Henneaux asymptotics.  Imposing Brown-Henneaux boundary conditions would excise both branches, leaving only the Einstein solutions.  Thus the approach taken in this paper may provide additional evidence of the equivalence of non-critical gravity and Einstein gravity.

\section*{Acknowledgements}
I would like to thank Steve Carlip for suggesting this project and for comments on early versions of this paper.  Additionally, I thank Eric Bergshoeff for helpful discussions.

\newpage
\appendix
 \renewcommand{\theequation}{A.\arabic{equation}}
  \setcounter{equation}{0}

\section{Equations of Motions in Fefferman-Graham Coordinates}
Here I present various formulae useful in solving the equations of motion.  The modified asymptotic expansion
\begin{eqnarray}
    ds^2 &=& dr^2 + \gamma_{ij}dx^idx^j \nonumber \\
    \gamma_{ij} &=& e^{2r}g_{ij} = e^{2r}\left(g^{(0)}_{ij} + e^{-nr}g^{(n)}_{ij} + e^{-2r}g^{(2)}_{ij} + \mathcal{O}(e^{-(n+2)r}) \right) \label{FGmetric}
\end{eqnarray}
is used to expand the equations of motion, and all terms are expanded to first order in $g^{(n)}$ and first order in $g^{(2)}$.  The generic expansion holds for $n\neq 1$ and $n$ not a positive even integer.  Note that when $n=1$, expressions quadratic in $g^{(1)}$ appear at the same order as terms linear in $g^{(2)}$, and this case must be treated separately.  For the Fefferman-Graham expansion when $n=1$, see \cite{Kwon:2012hr}.  The same is true for, say, $n=4$.  Then, terms quadratic in $g^{(2)}$ appear at the same order as linear $g^{(4)}$ terms.

For generic $n$, the Ricci tensor components are
\begin{eqnarray}
    R_{rr} &=& -2 + \frac{-n^2+2n}{2}e^{-nr}\text{Tr}g^{(n)} + \mathcal{O}^{-(n+2)r}\\
    R_{ri} &=& -\frac{n}{2}e^{-nr}\left(\nabla^k g^{(n)}_{ki} - \partial_i \text{Tr}g^{(n)}\right) -
        e^{-2r}\left(\nabla^k g^{(2)}_{ki} - \partial_i \text{Tr} g^{(2)} \right) + \mathcal{O}^{-(n+2)r}\\
    R_{ij} &=& -2e^{2r}g^{(0)}_{ij} + e^{(2-n)r}\left[\left(-\frac{1}{2}n^2+n-2\right)g^{(n)}_{ij} +
        \frac{n}{2}g^{(0)}_{ij}\text{Tr}g^{(n)}\right] + \nonumber \\
    & & e^0 \left[R^{(0)}_{ij} - 2g^{(2)}_{ij} + g^{(0)}_{ij}\text{Tr}g^{(2)} \right] + \mathcal{O}^{-nr}
\end{eqnarray}
The Ricci scalar is given by
\begin{equation}
    R = -6 + \left(3n-n^2\right)e^{-nr}\text{Tr}g^{(n)} + e^{-2r}\left[R^{(0)}+2\text{Tr}g^{(2)}\right] + \mathcal{O}^{-(n+2)r}
\end{equation}

The new tensor in the equations of motion consist of several pieces, and it is useful to expand each piece separately.\\
$R_{\mu \alpha \nu \beta}R^{\alpha \beta}$
\begin{eqnarray}
  R_{r \alpha r \beta}R^{\alpha \beta} &=& 4 + \left(\frac{3n^2}{2}-4n \right) e^{-nr} \text{Tr} g_{(n)}\\
  R_{r \alpha i \beta}R^{\alpha \beta} &=& \frac{n}{2} e^{-nr}\left[\nabla^k g^{(n)}_{ki} -
    \partial_i \text{Tr} g_{(n)} \right]\\
  R_{i \alpha j \beta}R^{\alpha \beta} &=& 4e^{2r}g^{(0)}_{ij} + e^{(2-n)r}\left[ \left( \frac{n^2}{2}
    -n + 4\right)g^{(n)}_{ij} + \left(n^2-\frac{7n}{2}\right)g^{(0)}_{ij} \text{Tr} g_{(n)} \right]
\end{eqnarray}

$RR_{\mu \nu}$
\begin{eqnarray}
  RR_{rr} &=& 12 + \left(5n^2 -12n\right) e^{-nr}\text{Tr}g_{(n)}\\
  RR_{ri} &=& 3n e^{-nr} \left[\nabla^k g^{(n)}_{ki} - \partial_i \text{Tr}g_{(n)} \right]\\
  RR_{ij} &=& 12 e^{2r} g^{(0)}_{ij} + e^{(2-n)r} \left[\left(3n^2 - 6n + 12\right)g^{(n)}_{ij}
    + \left(2n^2 - 9n \right)g^{(0)}_{ij}\text{Tr} g_{(n)}\right]
\end{eqnarray}

$R_{\alpha \beta}R^{\alpha \beta}g_{\mu \nu}$
\begin{align*}
  R_{\alpha \beta}R^{\alpha \beta}g_{r r} &= 12 + \left(4n^2 -12n \right) e^{-nr}\text{Tr} g_{(n)}\\
  R_{\alpha \beta}R^{\alpha \beta}g_{r i} &= 0\\
  R_{\alpha \beta}R^{\alpha \beta}g_{i j} &= 12 e^{2r}g^{(0)}_{ij} + e^{(2-n)r}\left[ 12g^{(n)}_{ij} +
    \left(4n^2 - 12n\right) g^{(0)}_{ij}\text{Tr}g_{(n)}\right]
\end{align*}

$R^2 g_{\mu \nu}$
\begin{align*}
  R^2 g_{rr} &= 36 + \left(12n^2 - 36n\right)e^{-nr}\text{Tr}g_{(n)}\\
  R^2 g_{ri} &= 0\\
  R^2 g_{ij} &= 36 e^{2r}g^{(0)}_{ij} + e^{(2-n)r}\left[36 g^{(n)}_{ij} + \left(12n^2 -36n\right)
    g^{(0)}_{ij} \text{Tr}g_{(n)} \right]
\end{align*}

$\Box R g_{\mu \nu}$
\begin{align*}
  \Box R g_{rr} &= -n^2 (n-2)(n-3) e^{-nr} \text{Tr}g_{(n)}\\
  \Box R g_{ri} &= 0\\
  \Box R g_{ij} &= -n^2 (n-2)(n-3) e^{(2-n)r}g^{(0)}_{ij}\text{Tr} g_{(n)}
\end{align*}

$\nabla_\mu \nabla_\nu R$
\begin{align*}
  \nabla_r \nabla_r R &= n^3(3-n)e^{-nr}\text{Tr}g_{(n)}\\
  \nabla_r \nabla_i R &= n(n-3)(n+1)e^{-nr}\partial_i \text{Tr}g_{(n)}\\
  \nabla_i \nabla_j R &= n^2(n-3)e^{(2-n)r}g^{(0)}_{ij}\text{Tr}g_{(n)}
\end{align*}

$\Box R_{\mu \nu}$
\begin{align*}
  \Box R_{rr} &= e^{-nr}\left(-\frac{n^4}{2} + 2n^3 - n^2\right)\text{Tr}g_{(n)}\\
  \Box R_{ri} &= e^{-nr}\left[\left(-\frac{1}{2}n^3+n^2+n\right)\nabla^k g^{(n)}_{ki} + \left(\frac{1}{2}n^3
    -n^2 - 2n\right) \partial_i \text{Tr} g_{(n)} \right]\\
  \Box R_{ij} &= e^{(2-n)r}\left[\left(-\frac{n^4}{2}+2n^3-n^2-2n\right)g^{(n)}_{ij} + \left(\frac{n^3}{2}
    -2n^2 + n\right)g^{(0)}_{ij}\text{Tr}g_{(n)}\right]
\end{align*}

The gauge independent equations for $g^{(n)}$ are given in the text.  For $g^{(2)}$, the equations of motion are
\begin{eqnarray}
  \{rr\}: & \frac{1}{2}\left(\sigma - \frac{1}{2m^2}\right)\left(R^{(0)} + 2\text{Tr}g^{(2)}\right) = 0 \\
  \{ri\}: & \left(\sigma + \frac{1}{2m^2}\right)\left(\partial_i \text{Tr}g^{(2)} - \bar{\nabla}^j g^{(2)}_{ji}\right) = 0
\end{eqnarray}

\section{Brown-York Stress Energy Tensor in Fefferman-Graham Coordinates}
Here I present various formulae useful in computation of the Brown-York stress energy tensor.  The exponent $n$ is taken to be in the range $1<n<2$.  In this range, all quantities need be expanded only to ``first order'' in $g^{(n)}$ and $g^{(2)}$ in order to find all finite and divergent pieces of the boundary stress energy tensor.

Using the metric (\ref{FGmetric}), the inverse metric is given by
\begin{equation}
    \gamma^{ij} = e^{-2r}\left(g_{(0)}^{ij} - e^{-nr}g_{(n)}^{ij}-e^{-2r}g_{(2)}^{ij} + \cdots \right)
\end{equation}
where indices are raised and lowered with the inverse metric.  The extrinsic curvature (indices raised and lowered) and trace is just
\begin{eqnarray}
    K_{ij} \equiv -\frac{1}{2} \partial_r \gamma_{ij} &=& -e^{2r}g^{(0)}_{ij} + \frac{n-2}{2}e^{(2-n)r}g^{(n)}_{ij} +
        \mathcal{O}(e^{-nr})\\
    K = K_{ij}\gamma^{ij} &=& -2 + \frac{n}{2}e^{-nr}\text{Tr} g^{(n)} + e^{-2r}\text{Tr}g^{(2)} + \cdots
\end{eqnarray}


The auxiliary tensor is proportional to the Schouten tensor, $m^2 f^{\mu \nu} = 2\left(R^{\mu \nu} - \frac{1}{4}Rg^{\mu \nu}\right)$ and is expanded as
\begin{eqnarray}
    m^2 f^{rr} &=& -1 + \frac{-n^2+n}{2}e^{-nr}\text{Tr}g^{(n)} - \frac{1}{2}e^{-2r}\left(R^{(0)}+2\text{Tr}g^{(2)}\right) + \cdots\\
    m^2 f^{ri} &=&-ne^{-(n+2)r}\left(\nabla_k g_{(n)}^{ki} - \partial^i \text{Tr} g^{(n)}\right)
        -2e^{-4r}\left(\nabla_k g_{(2)}^{ki} - \partial^i \text{Tr}g^{(2)}\right)\\
    m^2 \hat{f}^{ij} &=& -e^{-2r}g_{(0)}^{ij} + e^{-(n+2)r}\left[\left(-n^2+2n+1\right)g_{(n)}^{ij} + \frac{n^2-n}{2}g_{(0)}^{ij}
        \text{Tr} g^{(n)} \right]\\
    & & + e^{-4r}\left[R_{(0)}^{ij} - \frac{1}{4}R^{(0)}g_{(0)}^{ij} + \frac{1}{2}g_{(2)}^{ij} +
        \frac{1}{2}g_{(0)}^{ij}\text{Tr}g^{(2)} \right]\\
    m^2 \hat{f} &=& \gamma_{ij}f^{ij} = -2 + ne^{-nr}\text{Tr}g^{(n)} + e^{-2r}\left(R^{(0)} + 2\text{Tr}g^{(2)}\right) + \cdots
\end{eqnarray}

Some final pieces necessary for computation of the boundary stress tensor include
\begin{eqnarray}
    -\nabla^{(i}\hat{h}^{j)} &=& \mathcal{O}(e^{-(n+4)r})\\
    \frac{m^2}{2}\mathcal{D}_r \hat{f}^{ij} &=& e^{-2r}g_{(0)}^{ij} + e^{-(n+2)r}\left[ \frac{1}{2}(n^2-2n-1)(n+2)g_{(n)}^{ij} +
        \frac{1}{4}(-n^2+n)(n+2)g_{(0)}^{ij}\text{Tr}g^{(n)}\right]\nonumber\\
    & & + e^{-4r} \left[-4R_{(0)}^{ij} + R^{(0)}g_{(0)}^{ij} - 2g_{(2)}^{ij} - 2g_{(0)}^{ij}\text{Tr}g^{(2)}\right] \\
    -m^2 K^{(i}_k f^{j)k} &=& -e^{-2r}g_{(0)}^{ij} + e^{-(n+2)r}\left[ \frac{-2n^2+5n+2}{2}g_{(n)}^{ij} + \frac{n^2-n}{2}
        g_{(0)}^{ij}\text{Tr}g^{(n)} \right] \\
    & & +e^{-4r}\left[2R_{(0)}^{ij} - \frac{1}{2}R^{(0)}g_{(0)}^{ij} + 2g_{(2)}^{ij} + g_{(0)}^{ij}\text{Tr} g^{(2)} \right]\\
    \nabla_k \hat{h}^k &=& \mathcal{O}(e^{-(n+2)r})\\
    -\frac{m^2}{2}\mathcal{D}_r \hat{f} &=& \frac{1}{2}n^2 g_{(0)}^{ij}\text{Tr}g^{(n)} + e^{-2r}
        \left[R^{(0)} + 2\text{Tr}g^{(2)} \right]
\end{eqnarray}

In addition to some of the pieces above, possible counterterms (\ref{GenCT}) include
\begin{eqnarray}
    m^4\hat{f}^2 &=&4 - 4ne^{-nr}\text{Tr}g^{(n)} - 4e^{-2r}\left(R^{(0)} + 2\text{Tr}g^{(2)}\right) + \mathcal{O}(e^{-(n+2)r})\\
    m^4 f_{ij}f^{ij} &=& 2 -2ne^{-nr}\text{Tr}g^{(n)} + e^{-2r}\left(-2R^{(0)}-4\text{Tr}g^{(2)}\right)
\end{eqnarray}

All together, the Brown-York stress energy tensor with indices lowered is
\begin{eqnarray}
    8\pi G T^{\text{BY}}_{ij} &=& e^{2r}\left(\sigma + \frac{1}{2m^2}\right) g^{(0)}_{ij}\nonumber \\
    & & + e^{(2-n)r}\left[\frac{4\sigma m^2+2n\sigma m^2 + 2n^3 - 4n^2 + n + 2}{4m^2} g^{(n)}_{ij} +
        \frac{-2nm^2\sigma - n^3 + 2n^2 - 2n}{4m^2}g^{(0)}_{ij}\text{Tr}g^{(n)}\right]\nonumber \\
    & & + e^0 \left[2\left(\sigma + \frac{1}{2m^2}\right)g^{(2)}_{ij} - \frac{1}{4m^2}R^{(0)}g^{(0)}_{ij}
        - \left(\sigma + \frac{1}{m^2}\right)g^{(0)}_{ij}\text{Tr}g^{(2)} \right]
\end{eqnarray}

\newpage

\newpage

\end{document}